# Vertex Dependent Dynamic Response of a Connected Kagome Artificial Spin Ice

Ali Frotanpour, Justin Woods, Barry Farmer, Amrit P. Kaphle, Lance DeLong

Department of Physics and Astronomy, University of Kentucky, 505 Rose Street, Lexington, Kentucky 40506-0055, USA

**Abstract:**

We present the dynamic response of a ***connected*** Kagome artificial spin ice with emphasis on the effect of the vertex magnetization configuration on the mode characteristics. We use broadband ferromagnetic resonance (FMR) spectroscopy and micromagnetic simulations to identify and characterize resonant modes. We find the mode frequencies of elongated, single-domain film segments not only depend on the orientation of their easy-axis with respect to the applied magnetic field, but also depend on the vertex magnetization configuration, which suggests control over the FMR mode can be accomplished by altering the vertex magnetization. Moreover, we study differences between the vertex center mode (VCM) and the localized domain wall (LDW) mode. We show that the LDW mode acts as a signature of the domain wall (DW) nucleation process and the DW dynamics active during segment reversal events. The results show the VCM and LDW modes can be controlled using a field protocol, which has important implications for applications in magnonic and spintronic devices.

**Introduction:**

Artificial spin ice (ASI) is a mesoscopic, 2-D lattice composed of elongated, single-domain, ferromagnetic thin-film segments. ASI are topologically frustrated systems with behavior analogous to the observed atomic-scale frustration in rare-earth pyrochlore compounds for which the ground state of the system is not unique [1-2]. The submicron length scale of ASI lattices makes them suitable candidates for magnonic crystals and spintronic devices [3-5]. Moreover, ASI are actively investigated as a new class of nano-scale systems for which classical models can be applied to probe the physics of magnetic frustration. ASI lattice types recently under study include the honeycomb lattice [6-11], square lattice [12-15] and quasicrystals [16-18], which exhibit interesting long-range order, field-driven reversal, spontaneous switching, and dynamic response.

Elongated thin-film segments in ASI are dominated by long-range dipolar interactions (including demagnetizing fields). If the segments are connected at their end points, a relatively large vertex area is created; dipolar and shorter-range exchange interactions compete to determine the magnetization texture within and near the vertex, which strongly influences segment switching and DW dynamics [8,11]. For example, Bang et. al. [11] recently showed that connected three-fold segment clusters support strong FMR response in the vertex center when the applied field is aligned along the easy axis of a subset of the segments.

In this paper, we use FMR spectroscopy to experimentally study the influence of vertices on the FMR modes of a connected Kagome ASI. We analyze our broad-band (BB) FMR data using Object Oriented Micromagnetic Framework (OOMMF) simulations. Our findings improve our understanding of the relation between FMR modes in connected ASI and the dynamic response of DW localized within vertices, especially during reversal events.

Generally, the highest-frequency mode in an array of **disconnected**, single-domain thin-film segments with similar dimensions is resident on those segments whose magnetization makes the smallest angle with the applied field [8-14].

First, we demonstrate that vertices in **connected** Kagome ASI can be initialized so that the highest-frequency mode is resident in segments whose angle between their Ising segment magnetization and the applied field **is not** the smallest in the array.

Second, we found it useful to investigate DW modes residing within the vertices, and characterize their FMR modes before and after reversal events. We show deformations of vertex magnetizations induced by applied field strongly affect the slope, df/dH, of the frequency-field resonance curve, and DW located in vertices play a decisive role in the reversal of adjacent segments.

The results suggest FMR modes active in certain segments can be controlled via a specific field protocol that is suitable for application in magnonics. Characterization of the vertex DW modes is therefore essential for engineering spintronics applications that mainly rely on local excitation of DW.

**Methods:**

Experimental FMR spectroscopy was performed on Kagome ASI with connected vertices (see **Fig. 1**). Permalloy thin films were patterned on $SiO_2$ substrates using electron beam lithography followed by electron beam evaporation and lift-off. The width, length and thickness of the permalloy segments were 140 nm, 540 nm and 25 nm, respectively. We distinguish between three subsets of Ising segments designated A, B, and C (see **Fig. 1**) according to the angle of their easy axis with respect to the **x**-axis. Segments A, B and C correspond to $\varphi_A = 0°$, $\varphi_B = -60°$ and $\varphi_C = 60°$, respectively. The diameter of a single patterned Kagome ASI was about 40 micron, and a $4 \times 80$ array of these ASI was fabricated with a center-to-center distance of 100 microns.

Broadband FMR spectroscopy was used to probe the dynamic response of the magnetization. A microstripline of 12 mm length, 320 micron width and 100 micron substrate thickness was fabricated using standard photolithography techniques. The middle of the microstripline is composed of four strips of 20-micron width and 100-micron spacing. The sample was placed in a flip-chip geometry on the microstripline such that the ASI arrays lined up with the 20-micron strips. A vector network analyzer (VNA) was used to measure the transmission coefficient, $S_{12}$ in the frequency range, 3-14 GHz. The $S_{12}$ data measured at 3000 Oe were subtracted from the $S_{12}$ measured at each applied magnetic field to remove spurious background resonances.

We initialize the lattice by applying an Ising saturation field at zero angle ($\varphi = 0$) with respect to a reference axis formed by one of the Kagome lattice segments; we then perform broadband FMR measurements with the applied field oriented at 60° with respect to the reference axis; consequently, one can investigate the effects of the vertex magnetization texture on the dynamic response of most of the segments in the Kagome ASI.

The following field protocol was used to perform BB FMR measurements: Step 1: A magnetic field was applied along $\boldsymbol{\varphi} = 0$ to attain Ising saturation at H = 1000 Oe (see **Fig. 1**). Step 2: the magnetic field was turned off. Step 3: The magnetic field was swept from zero to 700 Oe in 10-Oe steps at $\boldsymbol{\varphi} = +60°$, and BB FMR measurements were performed at each applied field step. Consequently, the Ising saturation magnetizations of Segments A, B, and C, make angles of 60°, -120°, and 0° with respect to the ramping applied field, respectively. Therefore, as we increase the applied field from zero to 700 Oe with $\boldsymbol{\varphi} = +60°$, Segments B are expected to reverse at a specific field.

We used Object Oriented Micromagnetic Framework (OOMMF) to analyze our experimental results, using a 10-nm by 10-nm in-plane pixel area and a 25-nm permalloy thickness. The simulations assumed a saturation magnetization $M_s$ = 800 kA/m, and exchange stiffness $A = 1.3 \times 10^{-12}$ J/m [19]. In simulations, ASI are saturated in fields oriented along $\boldsymbol{\varphi} = 0°$, and then the magnetic field is swept from zero to 600 Oe with 5 Oe steps along $\boldsymbol{\varphi} = 60°$. The magnetization textures generated by OOMMF were used as input to simulate the FMR data: A 10-Oe pulse of magnetic field was applied perpendicular to the plane of the film with a duration of 20 ps. The magnetization vectors were recorded every 20 ps (i.e., 1000 times). The damping coefficient in the micromagnetic simulation for FMR was 0.008. The FMR absorption spectrum was found by applying a Fast Fourier Transform (FFT) on each pixel, and averaging over all pixels.

**Results and Discussion:**

**Figures 2 (a)** and **(b)** show the experimental and simulated FMR results. We observed a sudden change in the FMR spectrum at an applied field H = 260 Oe; this suggests reversals have occurred among the Segment B subgroup, since their remnant magnetization makes an angle of 120 degrees with respect to the applied field.

Five modes are observed in the experimental FMR spectra prior to Segment B reversal; and they are labeled $M_A$, $M_{B1}$, $M_{B2}$, $M_C$ and LDW (Localized Domain Wall). Mode $M_A$ has almost zero df/dH, whereas Modes $M_{B1}$ and $M_C$ are degenerate at zero field, but split at higher fields, according to positive and negative values of df/dH. Moreover, Modes $M_{B2}$ and LDW are observed at relatively low frequencies compared to modes $M_A$, $M_{B1}$, $M_C$. Corresponding modes found in simulations (shown in **Figs. 2 (c)** and **(d)**) are in good agreement with these experimental results.

The spatial distribution of FMR absorption (mode profiles) correlates well with particular segment types or vertices, as shown in **Fig. 3 (a)** for applied field H = 200 Oe and $\varphi$ = 60°. Mode $M_A$ corresponds to Segment A at H = 200 Oe. Mode $M_C$ ($M_{B1}$) corresponds to Segment C (B) and Mode $M_{B2}$ corresponds to a different mode for Segment B having more anti-nodal lines (note the parallel lines in the power map with strong absorption) than Mode $M_{B1}$.

### A. Vertex control of FMR modes

The FMR modes extant before Segment B reversal show two remarkable features. First, the frequency of Mode $M_A$ is greater than Mode $M_C$, even though the magnetization of Segment C is in the direction of the applied field, and the magnetization of Segment A makes an angle of 60° respect to the applied field. Second, df/dH of Mode $M_A$ is zero, whereas it is expected to be positive [8-12] since the segment is magnetized nearly parallel to the easy-axis. We now demonstrate that these features of Mode $M_A$ are correlated with the vertex magnetization configuration.

We define the vertex region as a triangle, as shown in **Fig. 4**: Each vertex is defined by two DW that separate it from two neighboring segments, and a central region with almost uniform magnetization that has the same direction as the axis of the third segment. The vertex can have any of six different magnetization states, depending on the angle, $\psi$, of the magnetization of the central region with respect to the applied field. The example shown in **Fig. 4** is a "-60° vertex state" ($\psi$ = -60°), which means the magnetization of the vertex central region makes an angle of -60° with respect to the applied field.

**Figure 5** shows the simulated magnetization texture of three nearest-neighbor segments for different applied fields. As can be seen, the magnetization of all vertex centers are aligned with the magnetization of Segment A, which means that all vertices are in the state, -60°. The bulk (inside the segment body) FMR mode is influenced by the magnitude and distribution of the demagnetization field inside the segment body. In particular, FMR results suggest the "-60° vertex state" causes a higher demagnetization field in Segment C compared to Segment A; consequently, the frequency of mode $M_A$ is highest compared to mode $M_B$ and $M_C$. In other words, Segment C has a shorter effective length (region of uniform magnetization) compared to Segment A, which is consistent with the fact that a higher demagnetization field obtains for shorter segments [21].

In summary, even though the magnetization of Segment C is in the direction of the applied field, it has a higher demagnetization field, which leads to the lower mode frequency compared to Segment A. At the same time, we must understand why the frequency of mode $M_A$ remains constant as the field increases (zero df/dH), which inevitably causes the bulk magnetization of Segment A to tilt slightly towards the hard axis, which involves a compensating increase in the demagnetization field of Segment A.

Remarkably, the DW between Segments A and B begin to move into the interior of Segment B as the applied field increases, but ***before there is a significant deformation of the vertex magnetizations***. After Segment B reversal, the "-60° vertex state" switches to a "0° vertex state" (note the highest frequency mode after reversal corresponds to Segment C). This is an example of the importance of vertex magnetization in reversal. Note that reversal of the Type B Segments occurs at 260 Oe in the experiment and at 440 Oe in the simulation, consistent with results of a variety of micromagnetic simulations that predict a higher reversal field than is observed [6-8].

After reversal, we detected four distinct modes with further increases in applied field, labeled as $N_{AB}$, $N_C$, LDW and VCM. All of these modes had a positive df/dH. The simulated mode profiles at 550 Oe are shown in **Fig. 3 (b)** and confirm that the highest frequency Mode $N_C$ corresponds to Segment C and the Mode $N_{AB}$ resides in Segments A and B. Therefore, the condition for a segment to have the highest frequency mode is $\psi = 0°$ in this regime of applied field.

### B. Vertex modes

Vertex modes are observed in a lower-frequency regime compared to segment bulk modes, as shown in **Fig. 3 (b)**. As discussed above, each vertex area is bordered by two DW whose motion and dynamic response are tightly correlated with the segment switching process in connected ASI. We now characterize the modes associated with DW and show how they depend on the vertex magnetization orientation (i.e., $\psi$).

LDW field dispersion is shown in **Fig. 2**; and the absorption profiles before Segment B reversal for H = 0 Oe and H = 200 Oe and H = 350 Oe are shown in **Figs. 3 (a), (c)** and **(d)**, respectively. Note, LDW modes generated in simulations are in the same frequency range and exhibit similar df/dH as in experiment. However, simulated LDW modes consist of multiple bands spanning a range of resonant frequencies between 2 GHz and 4 GHz. The frequency of LDW modes decreases slightly with increasing applied field. Moreover, the antinodal lines of LDW mode profiles change as the field increases. Based on the magnetization textures for the different applied fields (see **Fig. 5**), the DW which span Segments A and B (DW$_{AB}$) move toward the bulk of Segments B as the field increases (as described above). A comparison of the antinodal lines of the LDW mode profiles (**Figs. 3 (a), (c)** and **(d)**) with DW$_{AB}$ locations at different fields shows that the antinodal lines are in the same locations as the DW. After Segment B reversal, the frequencies of the LDW modes and df/dH > 0 change suddenly. The LDW mode profile after reversal is shown in **Fig. 3 (b)** for 550 Oe. These results show that the frequency and df/dH of LDW modes can be controlled by the vertex orientation with respect to the applied field, $\psi$.

We also studied the vertex center mode (VCM), whose existence depends on the direction of the easy axis of one of the segments matching the applied field direction [11].

However, our simulations show this is not a ***sufficient*** condition for the existence of a VCM. The VCM is labeled in **Figs. 2 (b)** and **(d),** and the mode profiles for H = 0 Oe, and 550 Oe are shown in **Figs. 3 (c)** and **(d).** We did not detect a VCM before reversal in FMR experiments; in fact, our simulation shows this mode exists only below 100 Oe with weak amplitude before reversal, and disappears afterward. After reversal, this mode is found for fields greater than 500 Oe, both in the simulation and experiment. A careful comparison of the magnetization configurations of the vertex center at different fields reveals the VCM exists whenever the ASI adopts the "$0^o$ vertex state".

**Conclusion:**

We have demonstrated control of FMR modes by using a field protocol to set particular vertex magnetizations. We showed that the angle between the segment's easy-axis and applied field does not necessarily determine the highest frequency FMR mode in ASI. Furthermore, we showed that df/dH depends not only on the angle between the magnetization of the segment and its easy axis, but also depends on the vertex magnetization. Consequently, the vertex magnetization state offers an important variable to control FMR modes in ASI by applying different field protocols. For example, by local manipulation of the segment and vertex configurations, we can alter the FMR mode frequencies of the segments. Moreover, we characterized the VCM and LDW modes to better understand the internal magnetization configuration of the vertex, and how it signals the DW mode behavior. These results are suitable for engineering applications of spintronic and magnonic devices. Moreover, we have verified a modified condition for the existence of the VCM mode.

**ACKNOWLEDGMENTS**

Research at the University of Kentucky was supported by the U.S. NSF Grant DMR-1506979, the UK Center for Advanced Materials, the UK Center for Computational Sciences, and the UK Center for Nanoscale Science and Engineering. Research at the Argonne National Laboratory, a

U.S. Department of Energy Office of Science User Facility, was supported under Contract No. DE-AC02-06CH11357.

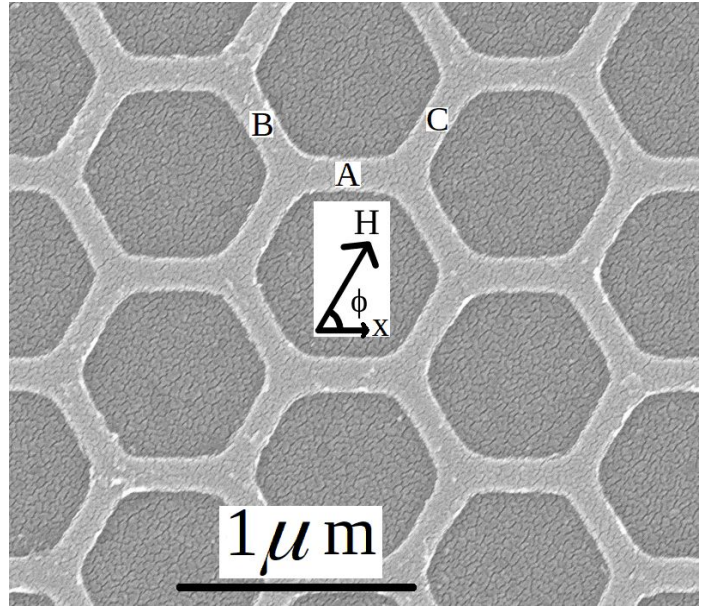

**Figure 1.** SEM image of a sample ASI.  The angle $\boldsymbol{\varphi}$ is defined with respect to the **x**-axis. Segments A, B, and C are labeled based on their values of $\boldsymbol{\varphi}_A = 0°$, $\boldsymbol{\varphi}_B = -60°$ and $\boldsymbol{\varphi}_C = 60°$, respectively.  Note the one-micron scale bar.

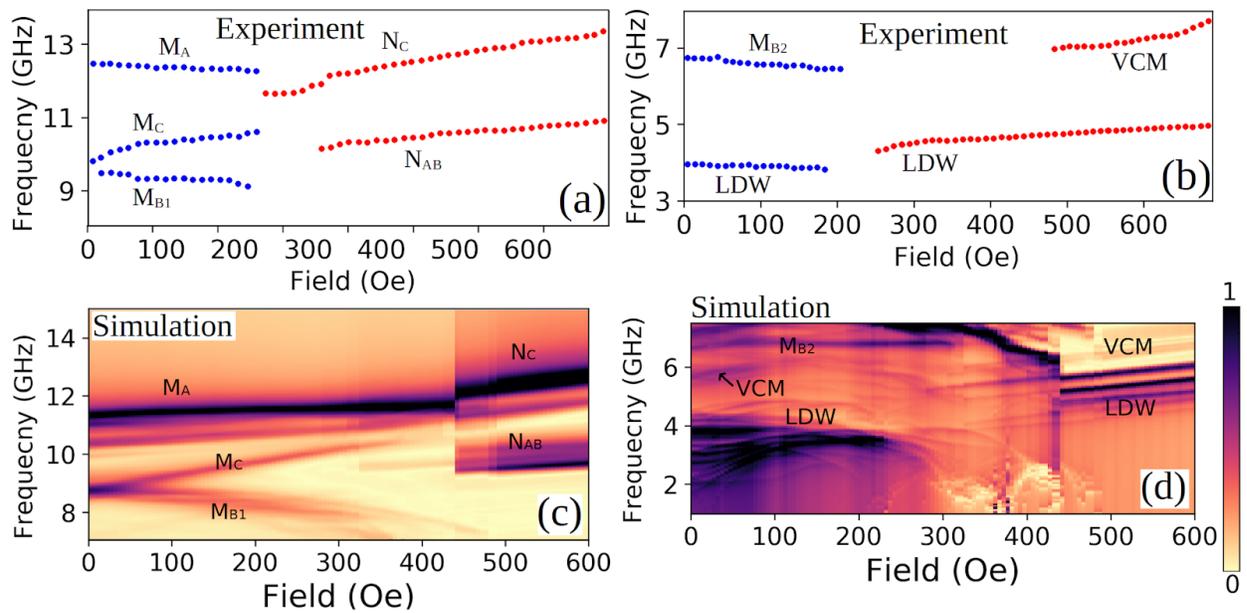

**Figure 2.** **(a)** and **(b)** show frequency-field graphs for experimental $S_{12}$ data for the 8-14 GHz and 3-8 GHz frequency bands, respectively. Panels **(c)** and **(d)** show simulated FMR absorption for frequencies 8-14 GHz and 3-8 GHz. The color scale corresponds to 1 for maximum signal absorption, and zero for the minimum absorption. Note the sudden change in the FMR spectrum at an applied field H = 260 Oe, which implicates reversals among the Segment B subgroup.

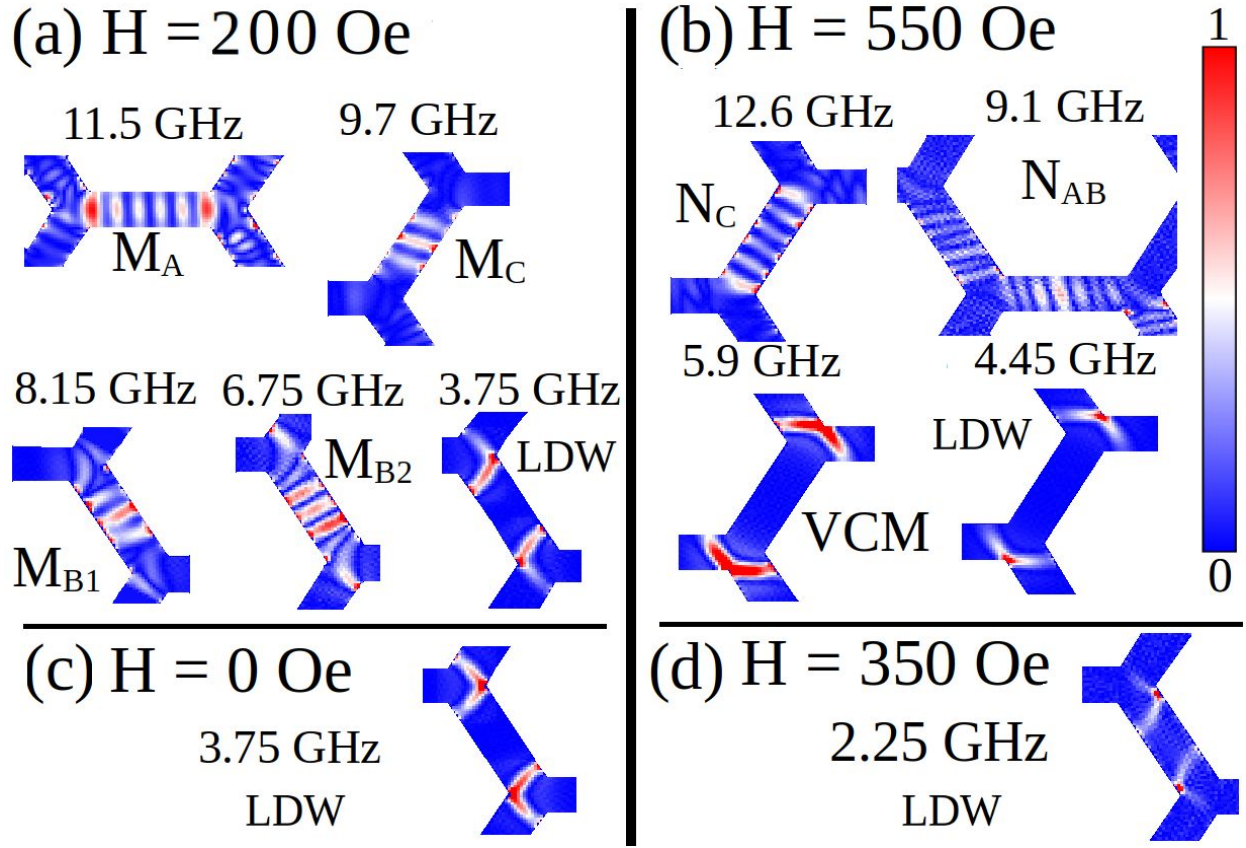

**Figure 3.** Mode profiles at different values of applied DC magnetic field H **(a)** 200 Oe, **(b)** 550 Oe, **(c)** 0 Oe, **(d)** 350 Oe. The color scale corresponds to 1 for maximum absorption and zero for minimum absorption. The applied field **H** is oriented at $\varphi_c = 60^o$ with respect to the horizontal **x**-direction. Segment B reversal occurs at 440 Oe in the simulations. A bulk FMR mode can be identified by antinodal lines in the body of the segments. LDW modes and VCM antinodal lines are visible in the vertex regions. LDW and VCM modes are rotated by $60^o$ after reversal (visible in (b) for 5.9 GHz and 4.45 GHz), which indicates the mode characteristics depend on the vertex magnetization state.

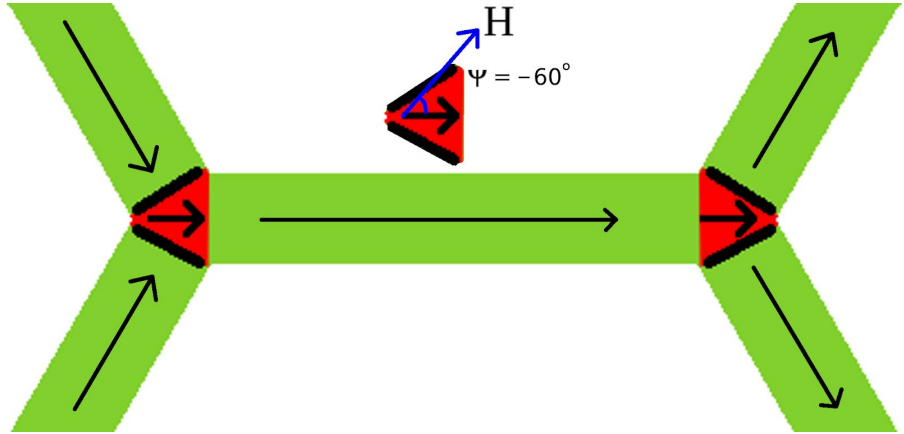

**Figure 4.** Vertex regions are highlighted as red triangles that connect neighboring segments. A central region with almost uniform magnetization is indicated by a black arrow inside a vertex, and two adjacent DW are shown as black lines. The angle $\psi$ between the vertex magnetization and the applied magnetic field $\boldsymbol{H}$ determines the state of the vertex. The example shown is a "-60° vertex state".

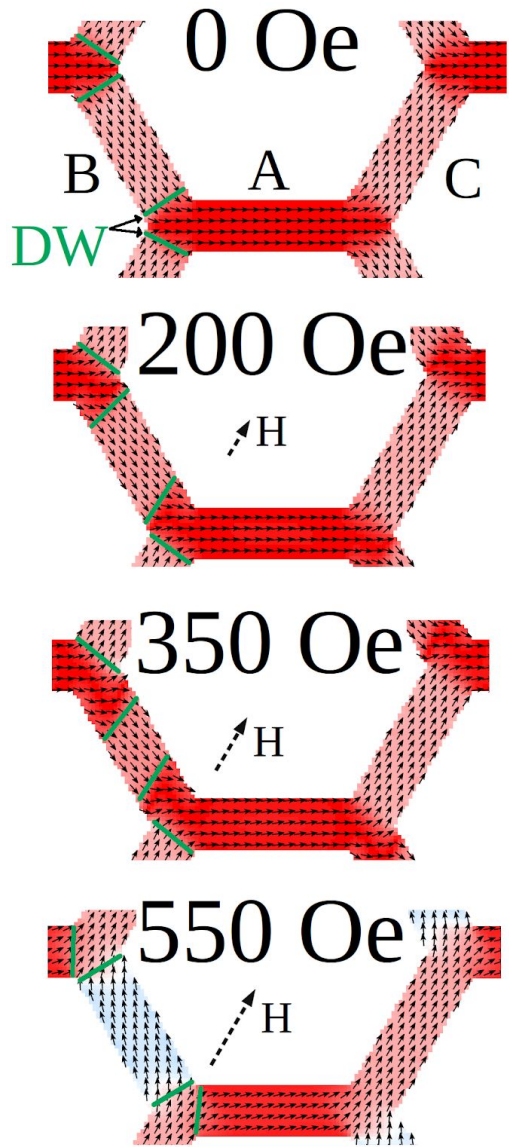

**Figure 5.** Magnetization configurations of Segment Types A, B, C and vertex textures for different applied fields. The DW shown by green lines separate vertex regions from Segments B and C for H = 0 Oe, 200 Oe, and 350 Oe. The vertex magnetization deforms as the field increases, and finally propagates, reversing Segment B. Note the stability of the vertex magnetization textures as the applied field is increased and the magnetization on the left segment is reversed.